\documentclass[preprint,10pt,5p,times,twocolumn]{elsarticle}
\usepackage{amsmath}
\usepackage{amsfonts}
\usepackage{amssymb}
\usepackage{amsthm}
\usepackage{graphicx}
\usepackage{color}
\journal{Physica A}
\begin{document}
\title{Spatiotemporal evolution in a (2+1)-dimensional chemotaxis model}
\author[SB]{S. Banerjee\corref{cor1}}
\ead{santo.banerjee@polito.it}
\author[APM]{A. P. Misra\fnref{fn1}}
\ead{apmisra@visva-bharati.ac.in}
\author[SB,LR]{L. Rondoni}
\ead{lamberto.rondoni@polito.it}
\cortext[cor1]{Corresponding author.}
\fntext[fn1]{Permanent address: Department of Mathematics, Visva-Bharati University, Santiniketan-731 235, India.}
\address[SB]{Dipartimento di Matematica, Politecnico di Torino, Corso Duca degli
Abruzzi 24, 10129 Torino, Italy}
\address[APM]{Department of Physics, Ume{\aa } University, SE-901 87 Ume{\aa }, Sweden}
\address[LR]{INFN, Sezione di Torino, Via P. Giuria 1, 10125 Torino, Italy}

\begin{abstract}
Simulations are performed to investigate the nonlinear dynamics of a (2+1)-dimensional  chemotaxis model of Keller-Segel (KS) type with a logistic growth term. Because of its ability to display auto-aggregation, the KS model has been widely used to simulate self-organization in many biological systems.  We show that the corresponding dynamics
may lead to a steady-state,  divergence in a finite time as well as   the formation of spatiotemporal irregular patterns. The latter, in particular, appear to be chaotic in part of the range of bounded solutions, as demonstrated by the analysis of wavelet power spectra. Steady states are achieved with sufficiently large values of the chemotactic
coefficient $(\chi)$ and/or with growth rates $r$ below a critical value $r_c$. For
$r > r_c$, the solutions of the differential equations of the model diverge in a finite
time. We also report on the pattern formation regime  for different values of $\chi$,
$r$ and   the diffusion coefficient $D$.
\end{abstract}
\maketitle
\section{Introduction}
The Keller and Segel (KS) model \cite{10} is one of the most extensively studied
chemotaxis models; it describes, for instance, the aggregation of cellular slime molds
driven by chemical attraction. Because it gives rise to  ``auto-aggregation'', the KS
model is used to describe many other phenomena including astrophysical ones \cite{7,8,9}.
Its mathematical properties, such as the existence of global solutions in time for
certain choices of parameters and the finite time divergence for other choices,
are understood to some extent \cite{chemotaxis-math-biology,chemotaxis-STC,chemotaxis-recent-work1}. In particular,
it is known that solutions of the one-dimensional case may diverge in a finite time
only if the chemoattractant does not diffuse, but in higher dimensional spaces
divergences are common \cite{chemotaxis-recent-work1}.
Also, Painter \textit{et al} have shown that the KS model allows pattern formation and
spatiotemporal chaos in one spatial dimension \cite{chemotaxis-STC}, but these features
have not been investigated in full detail in the (2+1)-dimensional case.

Therefore, in this paper, we consider the (2+1)-dimensional KS model with a
logistic growth term, which in dimensionless form  reads \cite{chemotaxis-math-biology,chemotaxis-STC}:
\begin{equation}
\frac{\partial u} {\partial t}=\nabla\left(D\nabla u-\chi u\nabla v\right)+ru(1-u),\label{1}
\end{equation}
\begin{equation}
\frac{\partial v} {\partial t}=\nabla^2v+u-v, \label{2}
\end{equation}
where $u(x,y,t)$ denotes the organism density and $v(x,y,t)$ describes the concentration
of the chemical signal around the
point $(x,y)$ at time $t$. The coefficients $D$ and $\chi$ respectively stand for the organism diffusion or motility  and the chemotactic sensitivity. The validity of Eqs. (\ref{1}) and (\ref{2}) in the framework of chemotaxis is supported by some experiments on the \textit{Escherichia coli (E.coli)} bacterium (see, e.g.\ Refs. \cite{11,6})  even if this model does not seem to reproduce all the observed chemotactic motions \cite{11}.

It is known that the solutions of Eqs. (\ref{1}) and (\ref{2})  with certain parameter choices 
reproduce an aggregation phenomenon which mimics chemotactic migration  along the directions
of gradients of self-produced chemicals. It remains to explore other regions of the
parameters space  in search of the parameter  values which lead to globally existing
solutions, the finite time divergence  as well as the formation of regular patterns. Furthermore, it is interesting to know whether chaotic patterns arise in regions of bounded solutions.

In the present work, we investigate these questions by means of numerical simulations,
and we show that the solutions of the KS equations in (2+1)-dimensions, with a logistic
growth term, diverge in a finite time  when the cellular growth rate $r$ exceeds a
critical value $r_c$.

A detailed discussion of different versions of the KS model and its applications to
chemotaxis can be found in the literature, e.g., in Refs. \cite{chemotaxis-math-biology,chemotaxis-STC,chemotaxis-recent-work1,chemotaxis-recent-work2,chemotaxis-recent-work3,chemotaxis-blow-up-prevention}.
Chemotaxis is a biological process by which organisms alter their movements or orientations 
in response to chemical gradients  and concerns, e.g.,  the  behaviors of 
\textit{E.Coli},  amoebae,   endothelial and neutrophil cells.
Furthermore, it is of interest in phenomena such as cancer cell metastasis, embryogenesis, angiogenesis, tissue homeostasis, wound healing, immune response, progression of diseases,
finding food, forming the multicellular body of protozoa  etc. \cite{E-coli-1,E-coli-2,E-coli-3}. In brief, chemotaxis is the process by which many organisms probe their environment
and determine their movements in response to what they have found.

\section{Nonlinear evolution}
Equations (\ref{1}) and (\ref{2}) have a nontrivial steady-state solution $(u,v)=(1,1)$.
A linear stability analysis \cite{chemotaxis-STC} through which different parameter regimes
can be found for the formation of spatiotemporal patterns shows that the corresponding modes
are unstable
for $\chi>\left(\sqrt{D}+\sqrt{r}\right)^2$ with wave numbers satisfying $k_1<k<k_2$ where
\begin{equation}
k^2_{1,2}=\frac{\chi-D-r}{2D}\mp\frac{1}{2D}\sqrt{\left(D+r-\chi\right)^2-4Dr}. \label{3}
\end{equation}
Thus, given a value of $\chi$, the stable (negative values) and the unstable (positive values) regions can be represented in the $(D,r)$ plane as in Fig. 1.
\begin{figure}
\includegraphics[width=4.0in,height=3in,trim=1.5in 0in 0in 0in]{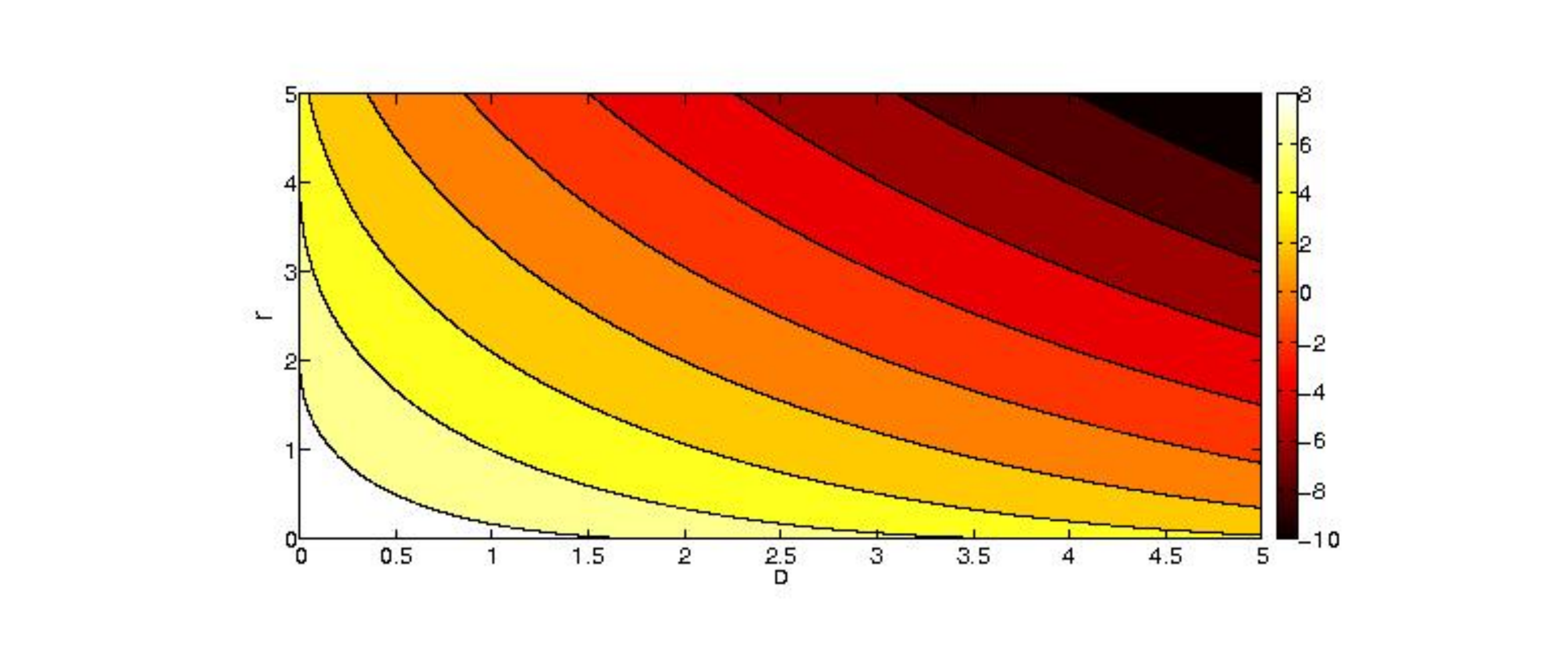} \caption{ (Color online) The instability regions  satisfying $\chi>\left(\sqrt{D}+\sqrt{r}\right)^2$ (positive values in the colorbar) are shown in the
$(D,r)$-plane.  }
\end{figure}
In an interval $[0, L]$ with homogeneous Neumann (zero-flux) boundary conditions we have
$k=n\pi/L$ with $n=0, 1, 2,...$, whilst for periodic boundary conditions we have $k=2n\pi/L$. We numerically investigate Eqs. (\ref{1}) and (\ref{2}) using zero-flux boundary conditions and with a time step $dt=0.001$, grid size $200\times 200$ in a simulation box of size $L_x\times L_y$
where $L_x=L_y=n\pi/k$. We explore the evolution of the density $u$ and of the chemical
concentration $v$ in a larger domain, with an initial condition $(u,v)=(1,1+f(\epsilon))$,
where $f(\epsilon)\lesssim0.01$ is a spatially varying random perturbation. Furthermore, we  perform a wavelet analysis of spectra to show that chaotic patterns may form in
some parameter region corresponding to bounded solutions. This is important for
the excitation of many modes in the nonlinear dynamics.

We note that for zero-flux boundary conditions on an interval of length $L_x$ or $L_y$, the smallest non-vanishing mode is the first, i.e.\ $n=1$ and $k=\pi/L_x=\pi/L_y$. Then,
using the condition of instability, for both bounded and unbounded solutions, one may
identify a parameter and a critical value below which patterns are regular, while they
turn irregular if the parameter is higher than this critical value. If the solution is
unbounded, it diverges in a finite time, as shown later. For example, fixing the parameter
values $D=0.1$ and $\chi=6$, we find that $r$ has a critical value $r_c=2$, below which
the wave pattern is regular or stable, and above which patterns exhibit irregular features. Furthermore, depending on the parameter values considered, there exist a critical value for
$k_0=k/n$, which delimits the range of chaotic pattern formation. The latter also depends
on the chemotactic nonlinearity for which the system exhibit finite solutions, as shown
in the following subsection.

To obtain steady-state oscillations, we consider the case with $k_0=0.07$ (i.e. $n\geq7$),
$D=0.1$, $\chi=6$, $k_1=0.45<k<k_2=6.9$ and two different values of $r$: $r=1$ (upper
panel of Fig. 2) and $r=1.5$ (lower panel of Fig.  2), which fall in the instability
region  (\textit{c.f.} Fig. 1). Figure 2 shows that  initially, the solution lies close to the
steady-state value $(1,1)$  before the organism kinetics dominates. Then, its density  or
the chemical concentration  drops to a steady-state profile determined by the balance
between the chemotactic nonlinear term and the dispersive diffusion term. We will later
see that the bounded solution may no longer exist and finite-time divergence may develop
if the value of $r$ is increased above $r=2$.
\begin{figure}
\includegraphics[width=4in,height=3in,trim=1in 0in 0in 0in]{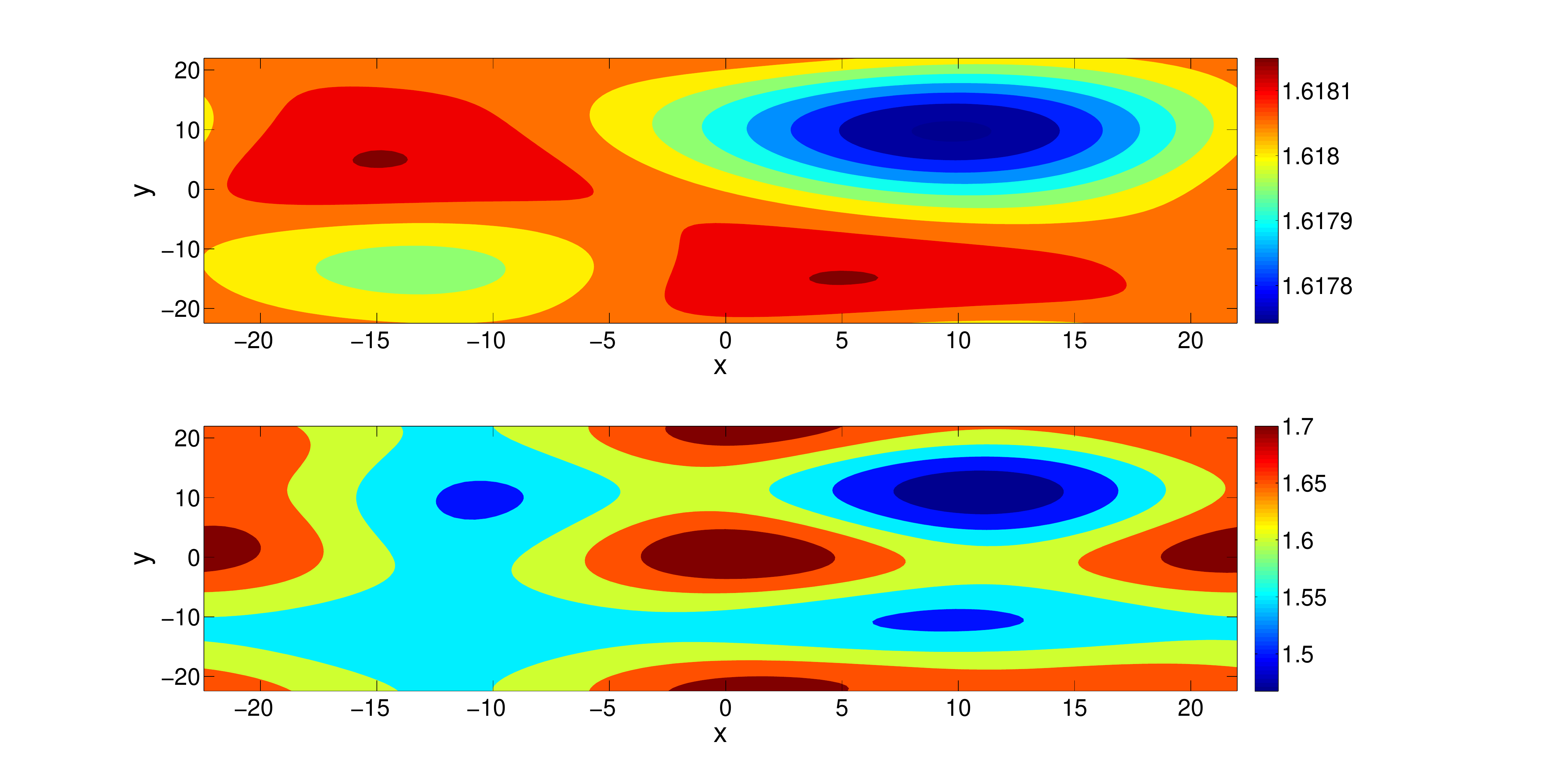} \caption{ (Color online) Numerical solutions of Eqs. (\ref{1}) and (\ref{2}) at $t=200$ for two different values of $r$: (a) $r=1$ and $r=1.5$ with a fixed $D=0.1$ and $\chi=6$. The solutions first increase around the steady-state $(1,1)$ and then drop to a more or less constant value for a longer time.}
\end{figure}
\subsection{Variation of the domain length}   As the value of $k_0$ is lowered, i.e.\ as $n$ increases in order that $k$   falls in the interval $(k_1,k_2)$, the domain length increases 
and the regular patterns disappear while irregular spatiotemporal structures arise  (see Fig. 3).
Basically, the pattern selection leads to the excitation of many more modes involved in
nonlinear interactions. As time advances, collision and fusion of patterns take place,
giving rise to higher harmonic modes (merging) and to the emergence of new modes. We
also note that increasing the domain size leads to higher concentrations of both $u$ and $v$
at a given time ($t=100$ in Fig.  3), and enhances the disorder of patterns  till a chaotic
evolution is established for $r\leq2$, the condition required for solutions to be bounded.
This is illustrated by the analysis of wavelet power spectra portrayed as in Fig.  4.
\begin{figure*}
\includegraphics[width=7in,height=4in,trim=0.0in 0in 0in 0in]{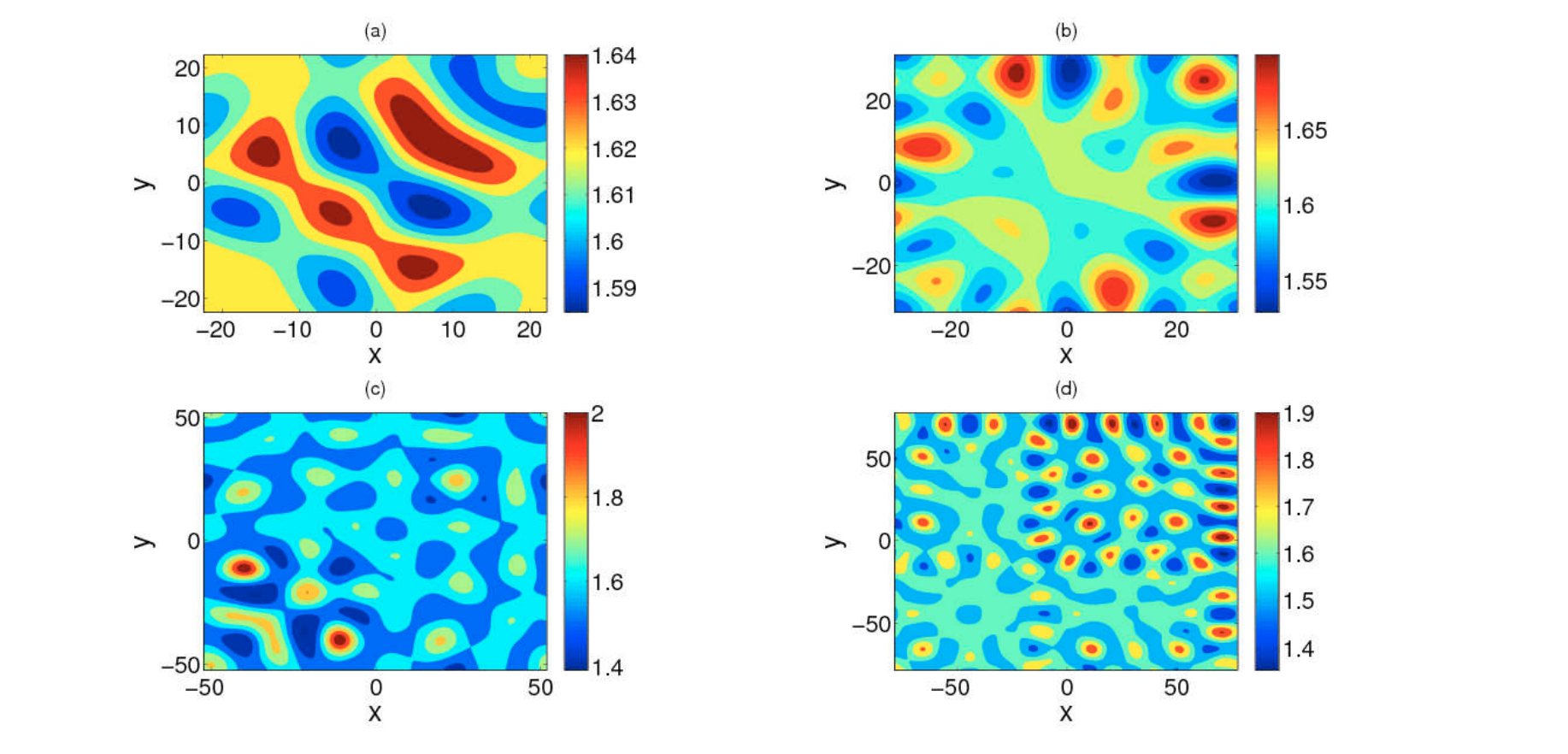} \caption{ (Color online) Numerical solutions of Eqs. (\ref{1}) and (\ref{2}) at $t=100$. Pattern formation is shown for different domain sizes: (a) $k_0=0.07$ (b) $k_0=0.05$  (c) $k_0=0.03$ and (d) $k_0=0.02$. The parameter values are $D=0.1$, $\chi=6$ and $r=2$.}
\end{figure*}
\begin{figure*}
\includegraphics[width=6.5in,height=3in,trim=0.0in 4.0in 0in 3in]{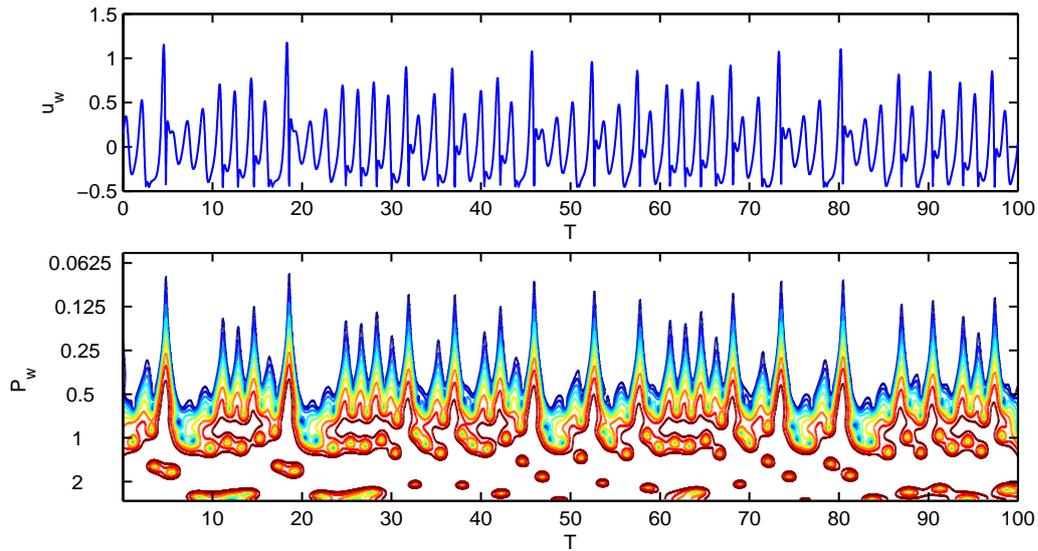}
\caption{ (Color online)   Normalized density wave $u_w$, rescaled by the
standard deviation
of the sampling data (upper panel) and wavelet power spectra $P_w$ (lower panel) as functions of the sampling time $T$
in logarithmic scale  corresponding to Fig. 4(d),   indicating chaotic wave trains.}
\end{figure*}
\subsection{Variation of the growth rate}
As $r$ exceeds its critical value $r_c=2$, the solutions quickly evolve and diverge in a
finite time (Fig. 5), where the density and the chemical concentration are reported
prior to the divergence time.  Figure 5  shows four different states at different times,
for $D=0.1$,  $\chi=6$ and  $r=4$. We find that the time of divergence depends on the
values of $r$: smaller $r$ implies longer time before the solution diverges.
For instance, the divergence time in Fig.  5 is $t\approx34$, whereas $D=0.1$,
$\chi=6$ and  $r=3$ imply a divergence at $t\approx52$.
Further increase of the growth rate $r$ leads to collision and fusion of patterns 
in a short time when the nonlinear logistic growth term dominates over the diffusion
and chemotactic terms. One can calculate the exact divergence times for other sets of
parameters.

It is to be noted that since the system is two-dimensional in space, the energy transferred
in the nonlinear interactions may not be due to the spatiotemporal chaos as in the previous
subsection, but could be primarily related to the wave collapse, i.e., to the divergence of solutions. However, this aspect needs further investigation, which will be subject of future work.
\begin{figure*}
\includegraphics[width=7in,height=4in,trim=0.0in 0in 0in 0in]{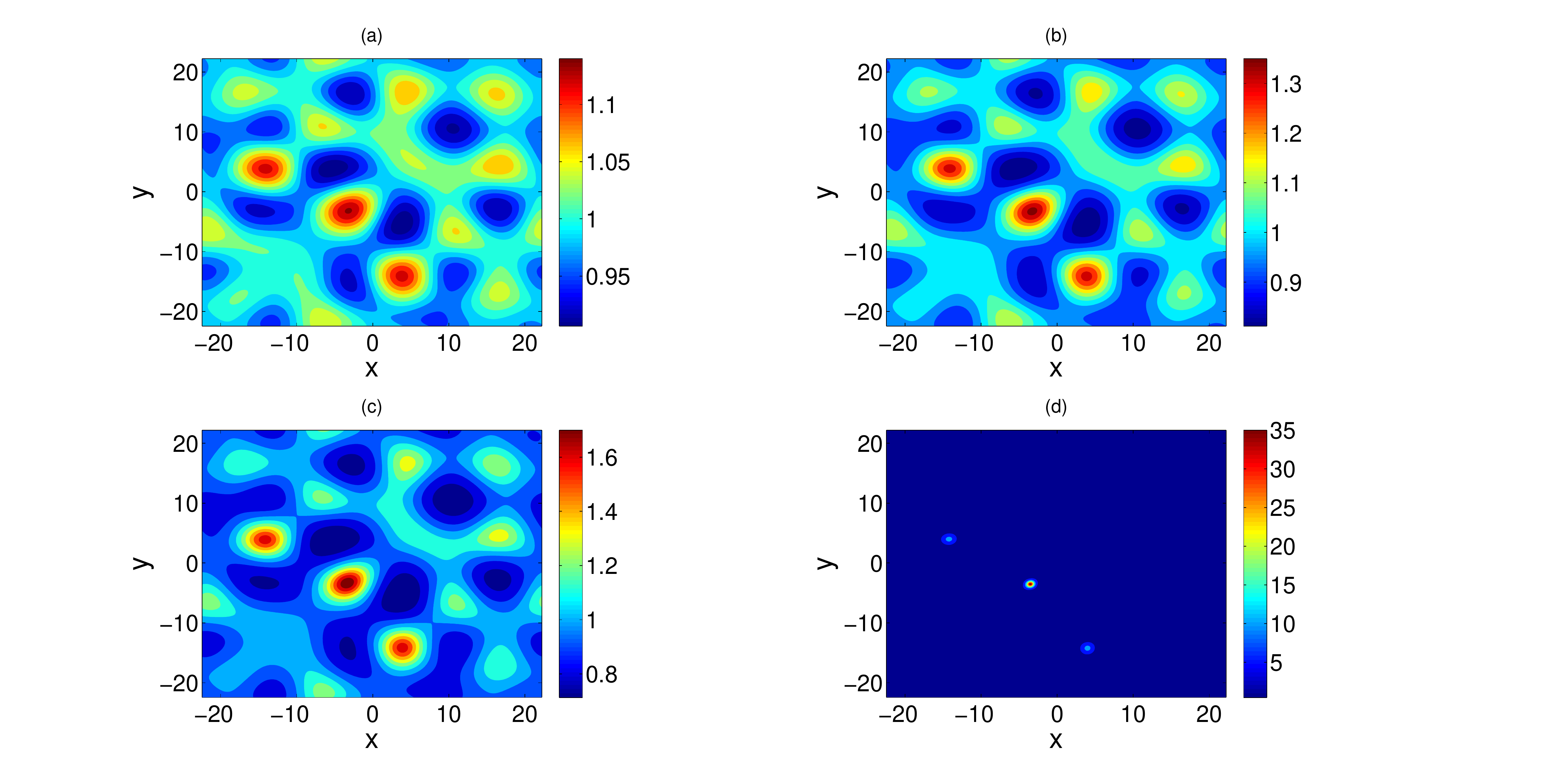} \caption{ (Color online) Numerical solutions of Eqs. (\ref{1}) and (\ref{2}). Pattern formation and solutions diverging in a finite time are shown at different times  before divergence:
(a) $t=25$, (b) $t=28$, (c) $t=30$ and (d) $t=34$.
The parameter values are $D=0.1$,  $\chi=6$ and $r=4$. }
\end{figure*}
\subsection{Combinations of $D$, $\chi$ and $r$} Next, we explore the pattern dynamics for a set of combinations of the system parameters, especially by increasing the diffusion as well as the chemotactic coefficient. Both parameters control various distinctive features. Comparing the
upper panel of Fig.  6  with Fig.  3(d), we note that increasing $\chi$ stabilizes oscillations
by reducing the values of $u$ and $v$. On the other hand, increasing the diffusion coefficient,
while keeping the other parameters fixed as in Fig. 3(d), one can observe chaotic patterns
similar to those of Fig. 3(d) (\textit{c.f.} lower panel of Fig. 6). This can be explained as follows.
As $\chi$ increases, with $D$ and $r$ fixed, the region of instability determined by the values
of $k$ increases, since its lower limit $k_1$ decreases while its upper limit $k_2$ increases.
In other words, $n$ may be smaller or fewer modes may be involved in the nonlinear dynamics
leading to chaos. On the other hand, when $D$ increases with other parameters
fixed, the instability region is strongly reduced, since the opposite happens with the lower
limit and upper limits for $k$, and $n$ has to be larger in order for $k$ to fall within the
instability region $k_1<k<k_2$. Thus, many more modes will participate in the collision and
fusion process, and as a result spatiotemporal chaotic patterns may likely arise.
\begin{figure*}
\includegraphics[width=7in,height=4in,trim=0.0in 0in 0in 0in]{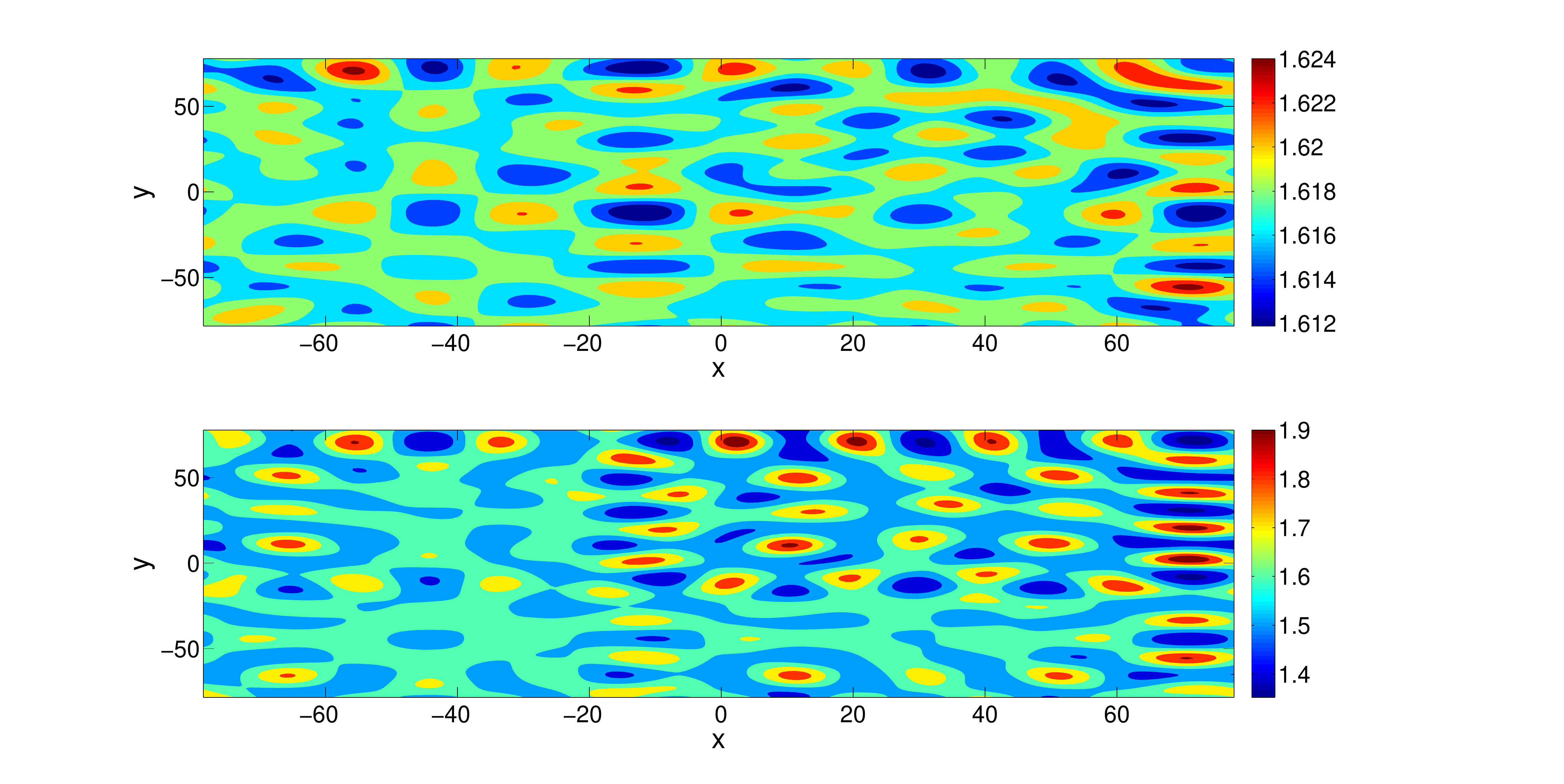} \caption{ (Color online) Pattern evolutions as solutions of Eqs. (\ref{1}) and (\ref{2}) for different combination of parameters: (a) $D=0.1$, $\chi=10$ and $r=2$ at $t=100$, (b) $D=1$, $\chi=6$ and $r=2$ at $t=100$.}
\end{figure*}
Figure 7 represents the contour plot of the wavelet power spectra, corresponding to Fig.  6,
plotted against a normalized sampling time $T$ (lower panel).
This shows that the system is chaotic for a long time interval. It is also interesting
to observe that the dynamics become more complex in nature at large values of $t$  [\textit{c.f.} Fig. 7(d)].
\begin{figure*}
\includegraphics[width=7in,height=3in,trim=0.0in 3.5in 0in 5in]{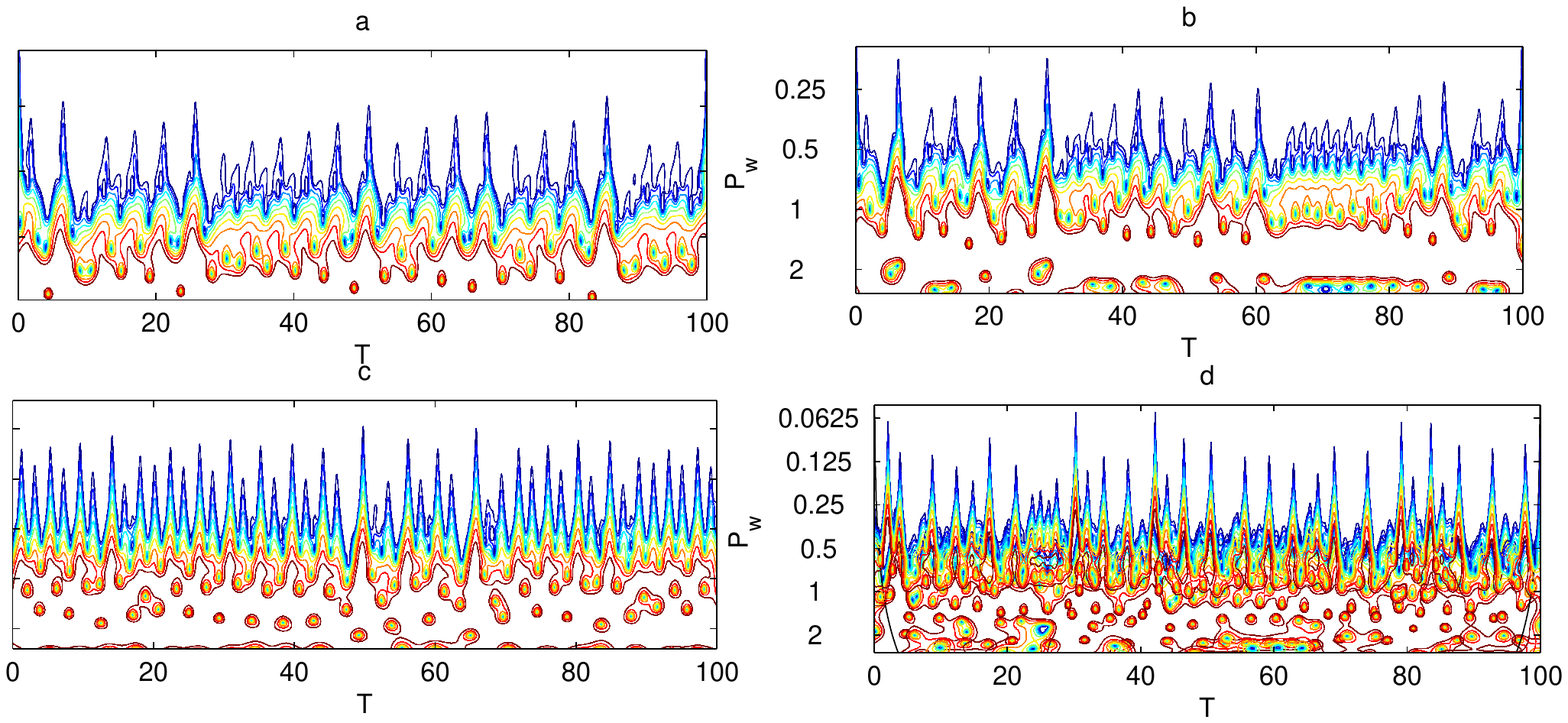}  
\caption{(Color online) Contour plot of the wavelet power spectra corresponding to Fig. 6 
(lower panel) for (a) t=100, (b) t=500, (c)=1000, (d) t=5000.}
\end{figure*}
\section{Conclusions}
We have explored the nonlinear dynamics of a (2+1)-dimensional chemotaxis model of KS-type. Specifically, by numerical simulations, we have focused on the existence of bounded solutions,  solutions which diverge in finite-time  as well as on the formation of different patterns including those arising from merging.

In the parameter range, where the solutions remain permanently bounded, the existence of spatiotemporal chaotic states is demonstrated by the analysis of wavelet power spectra.
These phenomena are important in the context of chemotactic migration but so far, to
the best of our knowledge, have not been explored in two-dimensional KS models.
We have shown that an increase of the chemotactic coefficient $(\lesssim10)$ and/ or
keeping the growth rate $(r\leq2)$, stable solutions around the steady-state are generated
after a long time $(t\gtrsim200)$. Furthermore, taking the growth rate beyond its critical
value leads to the finite time divergence of the solutions for the density and the chemical concentration. Concerning pattern formation and evolution in a large domain, suitable 
combinations of the parameters $\chi$, $D$ and $r$, e.g.\ $5\lesssim\chi\lesssim10$ 
and $1<r\lesssim2$, lead to bounded solutions which develop spatiotemporal chaos, whose 
complexity grows in time.
\section*{Acknowledgement}
A.P.M.   acknowledges support from the Kempe Foundtations, Sweden.
\bibliographystyle{elsarticle-num}

\begin{thebibliography}{50}
\bibitem {10} E.F. Keller, L.A. Segel,  J. Theor. Biol. {26} (1970)  399.
\bibitem{7} P. Biler,  Adv. Math. Sci. Appl. {9} (1999) 347.
\bibitem{8} P. Biler,  T. Nadzieja,  Colloq. Math. {66}  (1993) 131.
\bibitem{9} P. Biler,  T. Nadzieja, Report. Math. Phys. 52 (2003) 205.
\bibitem{chemotaxis-math-biology} T. Hillen,   K. J. Painter, J. Math. Biol. {58} (2009) 183.
\bibitem{chemotaxis-STC} K.J. Painter,  T. Hillen, Physica D {240} (2011) 363.
\bibitem{chemotaxis-recent-work1} Z. Wang, T. Hillen, Chaos {17} (2007) 037108.
\bibitem{11} M.P. Brenner, L. Levitov,  E.O. Budrene, Biophys. J. {74} (1995) 1677.
\bibitem{6} M.D. Betterton,  M.P. Brenner, Phys. Rev. E {64} (2001) 061904.
\bibitem{chemotaxis-recent-work2} J. Song, D. Kim, Phys. Rev. E {58} (2009) 183.
\bibitem{chemotaxis-recent-work3} Z.A. Wang, Math. Model. Nat. Phenom. {5}  (2010) 173.
\bibitem{chemotaxis-blow-up-prevention} Y.S. Choi, Z.A. Wang, J. Math. Anal. Appl. {362}  (2010) 553.
\bibitem{E-coli-1} H.C. Berg, D.A. Brown, Nature (London) {239} (1972) 500.
\bibitem{E-coli-2} J. Adler, Science {166} (1969) 1588.
\bibitem{E-coli-3} S.S. Willard, P.N. Devreotes, Eurro. J. Cell. Biol. {85}  (2006) 897.
\bibitem{1} W. Alt, J. Math. Biol. {9} (1980) 147.
\bibitem {53} H.G. Othmer, A. Stevens,  SIAM J. Appl. Math.  {57} (1997) 1044.
\bibitem {62} A. Stevens,  SIAM J. Appl. Math.  {61}  (2000) 183.








\end{thebibliography}

\end{document}